\documentclass[prl,twocolumn,superscriptaddress]{revtex4-1}
\usepackage{graphicx}
\usepackage{amsmath}
\usepackage{dcolumn}
\usepackage{color}
\usepackage{epstopdf}
\usepackage[caption=false]{subfig}

\newcommand\emf{\overline{\mbox{${\cal E}$}} {}}

\usepackage{float} 

\begin{document} 

\title{Dynamo-driven plasmoid formation from a current-sheet instability}

\author{F. Ebrahimi}  
\affiliation{Department of Astrophysical Sciences, and Princeton Plasma Physics Laboratory, Princeton University NJ, 08544}

\date{\today}
 
\begin{abstract}
Axisymmetric current-carrying plasmoids are formed in the presence of nonaxisymmetric fluctuations during nonlinear three-dimensional resistive MHD simulations in a global toroidal geometry. We utilize the helicity injection technique to form an initial poloidal flux in the presence of a toroidal guide field.  As helicity is injected, two types of current sheets are formed from 1) the oppositely directed field lines in the injector region (primary reconnecting current sheet), and 2) the  poloidal flux compression near the plasma edge (edge current sheet). 
We first find that nonaxisymmetic fluctuations arising from the current-sheet instability  isolated near the plasma edge have tearing parity but can nevertheless grow fast (on the poloidal Alfven time scale). These modes saturate by breaking up the current sheet. Second, for the first time a dynamo poloidal flux amplification is observed at the reconnetion site 
(in the region of the oppositely directed magnetic field). This fluctuation-induced  flux amplification increases the local Lundquist number, which then triggers a plasmoid instability and breaks the primary current sheet at the reconnection site. The plasmoids formation driven  by large-scale 
flux amplification, i.e. a large-scale dynamo, observed here has strong implications for astrophysical reconnection as well as fast reconnection events in laboratory plasmas.

\end{abstract}

\maketitle 

Burst-like occurrences of solar flares and magnetospheric substorms in nature, as well as sawtooth instabilities in magnetic-confinement laboratory plasmas, are believed to be initiated by magnetic reconnetion. Magnetic reconnection is a
major interplay for fundamental physical phenomena, such as particle acceleration and heating, magnetic-field
generation, and momentum transport. There is broad experimental or observational evidence of magnetic reconnection. (see for example \cite{zweibel09})
It is believed that there are two types of trigger mechanism for reconnection, forced or spontaneous magnetic reconnection. In forced magnetic reconnection, ~\cite{ji98,yamada2000} oppositely directed field lines are brought together as the result of  electromagnetic forces, or reconnection can be forced nonlinearly (so called driven reconnecton). Magnetic reconnection could also be spontaneous due to slow-growing global current-driven resistive tearing instabilities forming global islands or Alfvenic growing chains of plasmoids in a current sheet~\cite{shibata2001,loureiro07,bhattacharjee09}. In particular, spontaneous reconnection mediated by plasmoid instability has been shown to cause a fast reconnection rate in the resistive MHD
model.~\cite{loureiro2012,ebrahimi2015plasmoids}

Recently, using global MHD simulations, it has been shown that magnetic reconnection can also be induced via two mechanisms in magnetically confined fusion plasma of NSTX/NSTX-U during transient Coaxial Helicity Injection (CHI), the primary candidate for plasma current start-up. In forced reconnection, the oppositely directed field lines in the injector region are
forced to reconnect after the injector voltage is rapidly reduced to zero and  a stable Sweet-Parker (S-P) current sheet forms \cite{ebrahimi2013,ebrahimi2014}. In the second mechanism of
spontaneous reconnection, if helicity and plasma
are injected into the device at high Lundquist
number, the oppositely directed field lines in the
injector region spontaneously reconnect when
the elongated current sheet becomes MHD
unstable due to the plasmoid instability.\cite{ebrahimi2015plasmoids}
In this letter, by performing both 2-D and 3-D MHD simulations, we examine the effect of 
non-axisymmetric 3-D perturbations on the formation of reconnecting axisymmetric plasmoids. In particular, we explore whether the large-scale 
dynamo arising from 3-D fluctuations can induce axisymmetric 2-D plasmoid-mediated reconnection.  

In magnetically-dominated plasmas, reconnecting instabilities have also been shown to constitute a large-scale dynamo by converting one type of magnetic flux into
another. In the toroidal fusion configurations of Reversed-Field Pinches (RFPs) and spheromaks, the correlation between magnetic and flow fluctuations arising from tearing instabilities induces electromotive forces (emf) along the mean magnetic field.  The resulting emf modifies the background mean fields and contributes to current relaxation.  Current and momentum relaxation~\cite{choiprl,alexei2009} observed  during  reconnection events in the RFP were shown to be associated with both linearly unstable tearing modes as well as the nonlinearly-driven poloidally symmetric tearing modes. 

In this letter, we examine whether axisymmetric plasmoids could be driven/triggered by 3-D perturbations via a fluctuation-induced dynamo. This is inspired by the question of whether large-scale dynamo could have 
 a major contribution in triggering reconnecting plasmoids of solar observations,
 or could play a role in fast reconnection in fusion plasmas.
 To capture the key physics, simulations ought to be performed in three dimensions. We therefore employ a realistic three-dimensional toroidal global geometry  
during the helicity injection, where a \textit{reconnection site} exists near the helicity injection region. We then examine the interaction of non-axisymmetric magnetic fluctuations with the plasmoid instability. 
We find that 1) the edge currents that result from poloidal flux compression near the plasma edge could trigger edge current-sheet instabilities, 2) in 3-D simulations, when nonaxisymmetric fluctuations are included, axisymmetric plasmoids are formed, while for the same case in 2-D, 
 plasmoids were stable. In 3-D, the plasmoid-mediated reconnection occurs due to the flux amplification arising from the fluctuation-induced emf. The flux amplification near the reconnection site increases the local dimensionless Lundquist number  $S = L V_A/\eta$, so that the plasmoids instability can be triggered. Here,
the Alfven velocity  $V_A$ is based on the reconnecting 
magnetic field,  $L$ is the current sheet length, and $\eta$ is the magnetic diffusivity.

\begin{figure}
\includegraphics[]{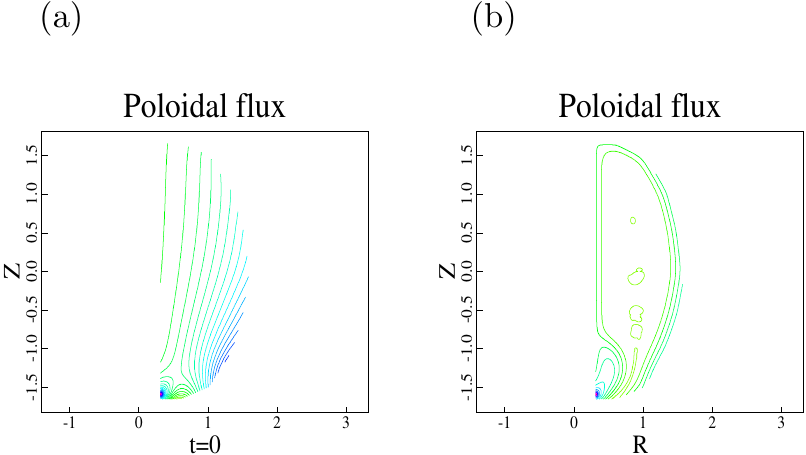}
\caption{(a) Initial poloidal flux, (b) subsequently injected poloidal flux fills up the global domain of 3-D simulations. Multiple plasmoids are formed.}
  \label{fig:fig1} 
\end{figure}

Simulations start with an initial poloidal field shown in Fig.~\ref{fig:fig1}(a).
In the presence of a 0.5~T toroidal field, a constant electric field is then applied to the poloidal flux footprints to drive current along 
the open field lines. Helicity is injected through the linkage of resulting toroidal flux with the poloidal injector flux. Figure \ref{fig:fig1}(b) shows the poloidal flux later in time as it is injected in the global domain.  Simulations are performed using the NIMROD code.~\cite{sovinec04} The discrete form of the equations
in NIMROD uses high-order finite elements to represent the poloidal plane (R-Z) and is pseudo-spectral
with FFTs for the periodic direction ($\phi$). We use a poloidal grid with 45 $\times$ 90 fifth-order finite elements in a global (R,Z) geometry, and toroidal mode of $n=0$ for 2-D simulations.   Perfectly conducting boundary condition is used, except at a narrow slot in the lower position (Z=-1.65m), which has a normal $E \times B$ flow where a constant-in-time electric field is applied at t=6ms.~\cite{ebrahimi2015plasmoids,ebrahimi16} The temporal resolution is
automatically adjusted to keep the 
flow CFL condition below a specified bound. However, to guarantee the convergence of the growth rates,  a time step as small as a nanosecond has been used  during the linear phase of the 3-D simulations.

Simulations presented here are resistive MHD simulations for a zero-pressure model (pressure  is not evolved in time). Magnetic diffusivity used in these simulations is $\eta = 5m^2/s$, and the kinematic viscosity is chosen to give a Pm =7.5 (Prandtl number = $\eta/\nu$). The Lundquist number in these simulations based on the reconnecting magnetic field can reach up to $S=2 \times 10^5$.

\begin{figure}
 \includegraphics[]{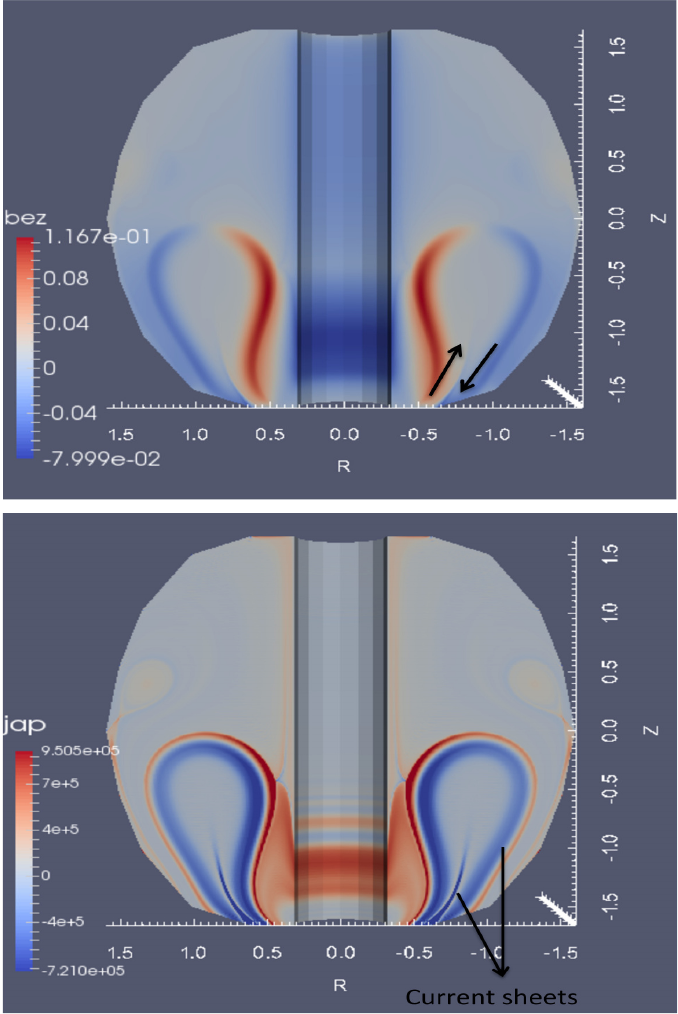}
 \caption{Two-dimensional global simulations in toroidal geometry at t=6.44s. Poloidal R-Z cuts of reconnecting magnetic field component $B_z$, the reconnection site is where the oppositely directed field lines in the injector region are marked (top), toroidal current density (bottom).}
  \label{fig:fig2} 
 \end{figure}

We first perform axisymmetric (with toroidal mode number $n=0$) time-dependent
 resistive MHD simulations. 
The vertical reconnecting magnetic field ($B_z$) and the associated toroidal current density ($J_{\phi}$) are shown during the early phase of the injection at t= 6.44ms, before the poloidal flux fills up the volume.
 As shown in Fig.~\ref{fig:fig2}(a), the vertical magnetic field changes sign in the injection region (the oppositely directed field lines shown by arrows), this region could thus provide a location for magnetic reconnetion. The poloidal cut of toroidal current density in Fig.~\ref{fig:fig2}(b) shows the formation of two types of current sheets during the helicity injection: 1) a current-sheet formed in the region of oppositely directed field lines, which is associated with the magnetic reconnection, 2) an edge current sheet, which is formed as the result of poloidal flux expansion  into the vacuum field. The edge current is formed (and induced) due to the radial magnetic compression of field lines (and  vertical expansion of poloidal flux) near the conducting vacuum vessel.  Both types of current sheets were also shown to exist in the previous axisymmetric simulations (see Fig.~2 in~\cite{ebrahimi2015plasmoids}). In the 2-D simulations shown in Fig.~\ref{fig:fig2}, the current sheet at the reconnection site is stable to plasmoids instability. No plasmoids are formed and reconnection does not occur during the helicity injection in  this particular 2-D simulation (Fig.~\ref{fig:fig2}). The absence of reconnecting plasmoids in this case has also been confirmed by the Poincare plot shown in Fig.~\ref{fig:fig4}(b). As will be discussed below, this is because the local Lundquist number is small and below the threshold value for plasmoids instability.

\begin{figure}
\includegraphics[width=3.2in,height=2.4in,angle=180]{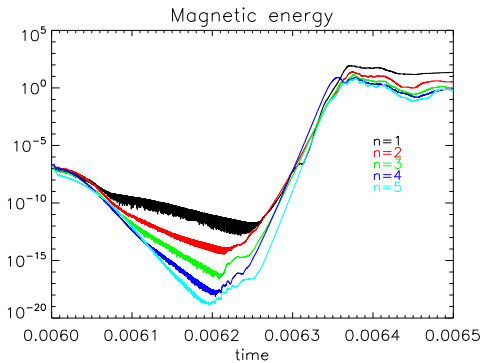}
\caption{Total non-axisymmetric modal energies vs. time in 3-D simulations, $\gamma \tau_{A (n=1)} = 0.16$, $\gamma \tau_{A (n=2)} = 0.18$, $\gamma \tau_{A (n=3)} = 0.2$,$\gamma \tau_{A( n=4)} = 0.23$, $\gamma \tau_{A (n=5)} = 0.26$.}
  \label{fig:fig3} 
\end{figure}

We next perform 3-D simulations by including toroidal mode numbers $n \neq 0$ (up to 22 toroidal modes are resolved). To examine the effect of non-axisymmetric perturbations, our 3-D simulations have been performed with \textit{exactly} the same parameters as the 2-D simulations shown above in Fig.~\ref{fig:fig2}. We find that non-axisymmetric modes start to grow during the helicity injection, as 
seen from the total magnetic energy evolution in Fig.\ref{fig:fig3}. These modes are edge current-sheet instabilities with high poloidal mode numbers m (high $k_z$), and are driven due to a large edge-localized current density. The high-n mode numbers saturate at much lower amplitudes.
As the poloidal flux is 
expanded into the volume, the edge current sheet gets further elongated. That means a local Lundquist number based on the length of the current-sheet could transiently increase during the injection (as well as the associated growth rates). As an evolving edge current sheet \textit{cannot} be treated as a static equilibrium~\cite{2016luca}, a true linear phase with a static equilibrium does not exist here. We therefore argue that it is more physically correct to calculate the growth rate of these modes during the early linear phase of the nonlinear evolution.

These modes localized in the edge current sheet have fast growth rates. The calculated early-phase growth rates are given in the caption of Fig.~\ref{fig:fig3}. Here, the relevant time scale is the poloidal  Alfven time, which is calculated based on the reconnecting poloidal magnetic field in the edge current sheet (see Fig.~2, $B_z$ =0.1T at around R=0.6m, with a relevant length of L=1m). These modes grow fast, on the poloidal Alfven time scales, but they have tearing-parity structures, i.e. even $B_r$ and odd $V_r$  in the layer.  The toroidal current density at a time after the saturation is shown in Fig.~\ref{fig:fig4}. As is seen, these high-$k_z$ edge modes break the background edge current density similarly to an axisymmetric current-sheet-driven plasmoid instability (see Fig.~2 
in~\cite{ebrahimi2015plasmoids}). These modes therefore saturate by modifying and relaxing the current sheet, in this case the edge current sheet.

\begin{figure}[!b]
\includegraphics[]{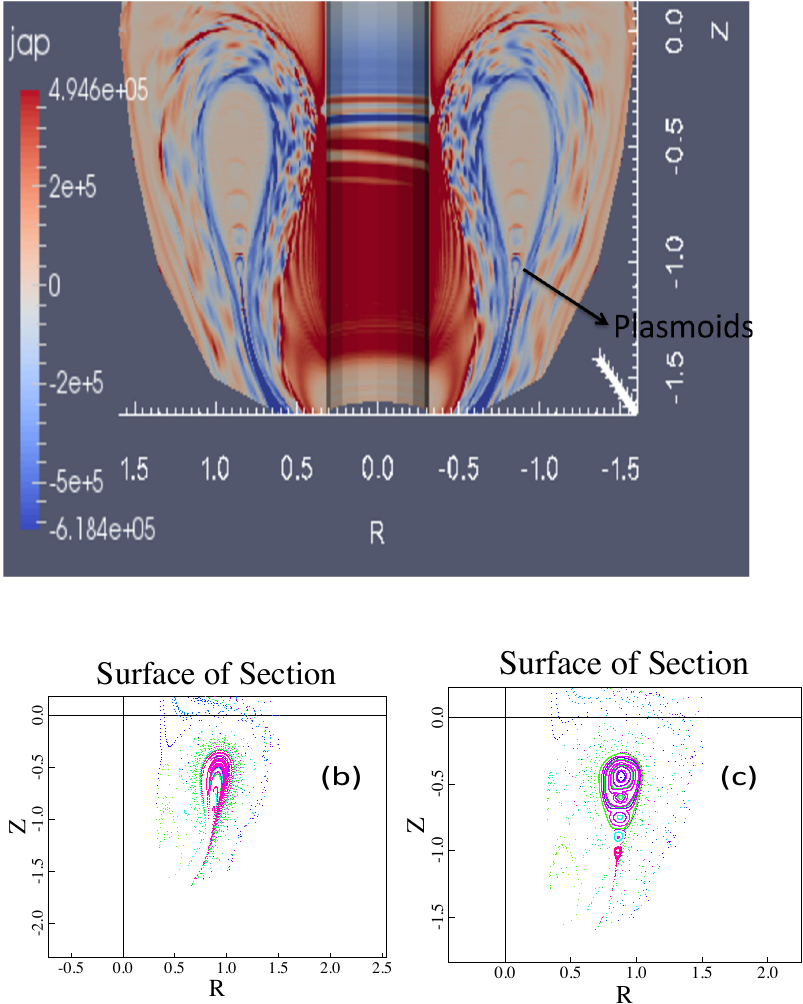}
\caption{Poloidal cut of toroidal current density $J_{phi}$ (top) in 3-D. Poincare plots, 
the intersections of a field line with a poloidal plane, as the field line is followed around
the torus (bottom) for (b) (2-D), no plasmoid (and no X-point) formed, field lines remain open and (c) (3-D) at least 5 plasmoids have formed.
}
  \label{fig:fig4} 
\end{figure}

\begin{figure}
\includegraphics[]{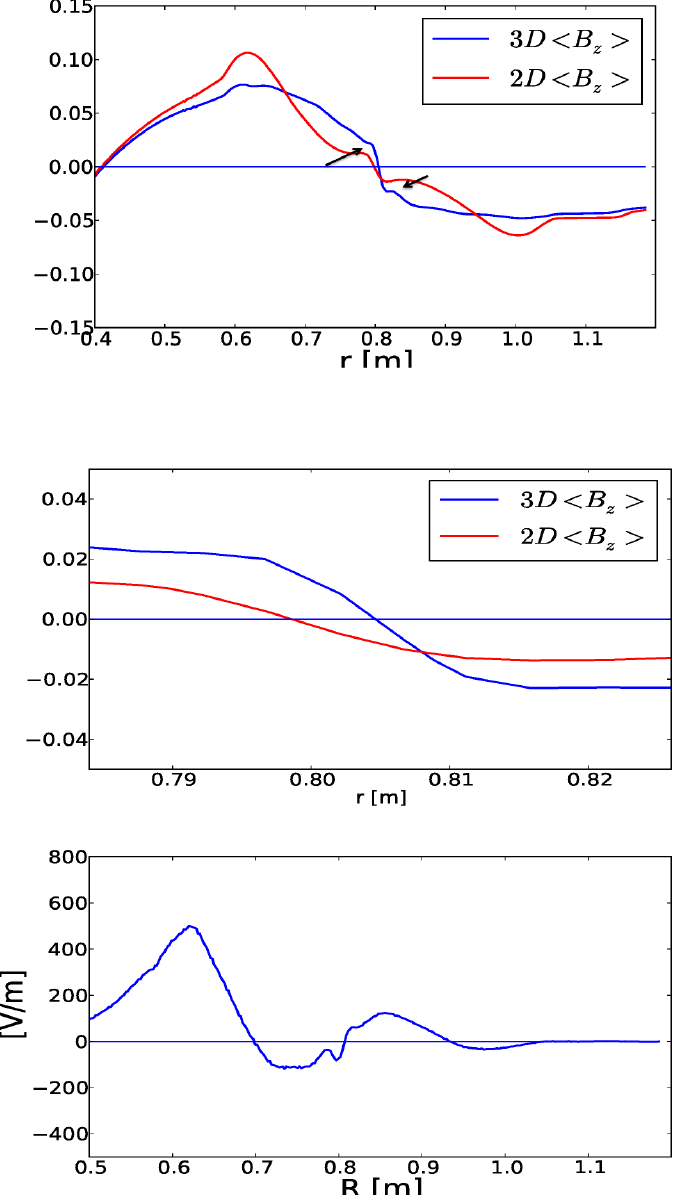}
\caption{Arrows show large-scale dynamo flux amplification/generation in 3-D. The radial profiles of the reconnecting component of the magnetic field $B_z$ at Z = -1.29m show amplification of the large-scale field in 3-D over the full radial extent (top), zoom in around $R \sim 0.8$ at the reconnecting site (middle). Mean toroidal component of emf, $\emf_{\phi} = <v \times b>_{\phi}$ 
(bottom).}
  \label{fig:fig5} 
\end{figure}

In addition to edge-localized modes arising from the current-sheet instabilities, the formation of axisymmetric plasmoids is also shown in Fig.~\ref{fig:fig4}.  Current-carrying plasmoids that are shown to be stable in a particular 2-D simulations in Fig.~\ref{fig:fig2}, are unstable in 3-D simulations in the presence of nonaxisymmetic toroidal modes. This can be explained by differences in the magnitude and the structure of the reconnecting field $B_z$ at the reconnection site in 2-D and 3-D, respectively, shown in Fig.~\ref{fig:fig5}. In 2-D simulations,  using the reconnection field at the reconnection site $B_z$ =0.012T (Fig.~\ref{fig:fig5}), the local Lundquist number is about $S \sim 5500$, which is below a threshold value of about $10^4$ for the plasmoids instability. However, as seen in Fig.~\ref{fig:fig5}, at the reconnection site, the vertical magnetic field in 3-D is significantly increased and showing a Harris-sheet type profile with a magnitude of  $B_z=$ 0.025T. This enhancement of reconnecting field in 3-D causes  the local Lundquist number to increase to a value of about $S \sim 15000$, above the plasmoid instability threshold.  The Poincare plots shown in Fig.~\ref{fig:fig4}(b,c) also confirm the formation of several plasmoids in the 3-D case, but no reconnection occurs in the 2-D case and the field lines remain open (Fig.~\ref{fig:fig4}(b)).

The poloidal-flux amplification around the reconnection site is only due to the presence of  magnetic fluctuations in 3-D. The edge fluctuation-induced emfs relax (and broaden) the edge current-sheet (seen from the edge current density, $J_{\phi}$, in Fig.~\ref{fig:fig4}) as well as the vertical magnetic field, $B_z$, shown in Fig.~\ref{fig:fig5} (in particular at R=0.8 and R=1.0 on both sides of the primary current sheet). This modification and broadening of $B_z$ around both sides of the primary reconnecting current sheet results in flux accumulation and formation of a Harris-sheet type profile and a subsequent plasmoid instability at the reconnection site.    From Faraday's equation, the flux amplification could be explained by the fluctuation-induced emf. Figure~\ref{fig:fig5} shows the mean ($n=0$) component of $\emf_{\phi} = <v \times b>_{\phi}$ 
at a vertical position, where a good correlation with the field modification is seen. The 2-D contribution of the $n=0$ component 
to the evolving $v \times b$ term has been subtracted. As is seen, the mean emf is bidirectional around $R=0.8m$ at the reconnection site, explaining the flux amplification resulting in a Harris-type profile in 3-D (Fig.~\ref{fig:fig5}b). We have also performed simulations with only one toroidal mode, and found that a single toroidal $n=1$ mode is sufficient to trigger plasmoids in 3-D. As the edge current sheet evolves via helicity injection,  nonaxisymmetric $n=1$ grows, but there could be nonlinear interaction between different poloidal mode numbers, $k_{z1}$ and $k_{z2}$, to produce an $n=0$ component to give a mean emf. This is similar to $m=0$ tearing modes in RFPs, which are nonlinearly driven from two $m=1$ modes with different toroidal mode numbers (of for example $n=6, 7$).~\cite{ebrahimithesis,choiprl} Such a comparison with further analysis of surface-averaged emf terms remains for future research.  

In summary, for the first time we have shown that 3-D magnetic fluctuations can cause flux amplification to trigger axisymmetric reconnecting plasmoids formation at the reconnection site. This is 
confirmed by performing both 2-D and 3-D simulations during helicity injection in a toroidal plasma. The 3-D effects are carefully compared with the 2-D simulation around the injection region
where oppositely-directly injected field lines could come together to reconnect. It is shown for stable current sheet in 2-D  that the spontaneous plasmoid reconnection could still occur due to a dynamo flux amplification from  nonaxisymmetric fluctuations. The 3-D magnetic reconnection observed here is important  for plasmoid-mediated reconnection as it could explain fast reconnection in astrophysical and laboratory plasmas (see for example \cite{ji2011}). 
The 3-D fluctuations could enhance the plasmoids formations at any Lundquist number S.  Here, we have demonstrated the 
minimum onset condition for the plasmoids formation in the presence of non-axisymmetric fluctuations, when the primary current sheet has been stable in 2-D. Our simulations (not shown) at much higher S, when the current sheet is unstable in 2-D, also show that the 3-D fluctuations could still enhance the local S to further increase the number of plasmoids. We therefore believe that these results also show that in the case of solar flares, 3-D nonaxisymmetric perturbations could generate large-scale magnetic field at the reconnection site to increase the local Lundquist number and then trigger/enhance axisymmetric plasmoids formation. 

Finally, the implication of these results for fusion plasmas will be reported in a separate paper. We find that the nonaxisymmetric edge modes are similar in nature to the axisymmetric current-sheet-driven plasmoid instability  reported in \cite{ebrahimi2015plasmoids}. The nonaxisymmetric modes observed here are edge-localized modes and further scaling of $S$ and $J_{\|}/B $ will also be reported. The current-sheet density  isolated near the edge may also be relevant to disruption in tokamaks.    We believe that an interesting characteristic of the edge current-sheet 
modes (concentrated in the open field lines) is that 
they exhibit peeling-type mode structures (Fig.~\ref{fig:fig4}(a)), similar to the peeling-mode experimental observation in spherical tokamaks.~\cite{2011peeling} Our preliminary scaling with $J_{\|}/B$ in fact shows the characteristic of edge peeling modes,~\cite{1998peeling} but with tearing parity.~\cite{2005peeling} Three-dimensional simulations  at NSTX-U relevant parameters, with the implication of edge current-sheet instability, and their effect on the global reconnection and flux closure in transient CHI will be reported in a future paper.

 We acknowledge Prof.~S. Prager, and Dr.~R. Raman for their thoughtful comments on this Manuscript. This work was supported by DOE grants DE-SC0010565, DE-AC02-09CHI1466, and DE-SC0012467.

\end{document}